\documentclass[useAMS,usenatbib,usegraphicx]{mn2e}
\newcommand{\kms}{\, \rm{km}\,  \rm{s}^{-1}}

\def\ltsima{$\; \buildrel < \over \sim \;$}
\def\lsim{\lower.5ex\hbox{\ltsima}}
\def\Msunh{\mbox{$h^{-1}$M$_\odot$}}

\def\Mpch{\mbox{$h^{-1}$Mpc}}
\def\rhovoid{\mbox{$\rho_{\rm void}$}}
\def\Mpc{{\rm Mpc}}

\def\deg{\ifmmode{^\circ}\else{$^\circ$}\fi}
\def\hGpc{\ifmmode{h^{-1}{\rm Gpc}}\else{$h^{-1}{\rm Gpc}$}\fi}
\def\hkpc{\ifmmode{h^{-1}{\rm kpc}}\else{$h^{-1}{\rm kpc}$}\fi}
\def\hMpc{\ifmmode{h^{-1}{\rm Mpc}}\else{$h^{-1}{\rm Mpc}$}\fi}
\def\hMsun{\ifmmode{h^{-1}M_\odot}\else{$h^{-1}M_\odot$}\fi}

\def\kms{{{\rm km}/{\rm s}}}

\def\muK{\ifmmode{\mu{\rm{K}}}\else{$\mu$K}\fi}
\def\mum{\ifmmode{\mu{\rm{m}}}\else{$\mu$m}\fi}

\newcommand{\aj}{{AJ}}
\newcommand{\apj}{{ApJ}}
\newcommand{\apjl}{{ApJ}}
\newcommand{\apjs}{{ApJS}}
\newcommand{\mnras}{{MNRAS}}
\newcommand{\nat}{{Nature}}
\newcommand{\aap}{A\&A}

\newcommand{\araa}{ARA\&A}
\newcommand{\prd}{Phys. Rev. D}

\title{The structure of voids}

\author[Stefan Gottl\"ober et al.]
{Stefan Gottl\"ober$^{1}$\thanks{E-mail:
sgottloeber@aip.de}
Ewa L. {\L}okas$^{2}$,
Anatoly Klypin$^{3}$, and
Yehuda Hoffman$^{4}$
\\
$^1$
Astrophysikalisches Institut Potsdam,
An der Sternwarte 16,
14482 Potsdam, Germany.
\\
$^2$
Nicolaus Copernicus Astronomical Center, Bartycka 18, 00-716 Warsaw,
Poland
\\
$^3$
Astronomy Department,
New Mexico State University, Dept.\ 4500,
Las Cruces, NM 88003, USA
\\
$^4$
Racah Institute of Physics, Hebrew University, Jerusalem 91904, Israel
}

\begin{document}

\maketitle

\begin{abstract}

Using high resolution $N$-body simulations we address the problem of
emptiness of giant $\sim 20\hMpc$--diameter voids found in the distribution of
bright galaxies. Are the voids filled by dwarf galaxies? Do
cosmological models predict too many small dark matter haloes inside the
voids? Can the problems of cosmological models on small scales be
addressed by studying the abundance of dwarf galaxies inside voids?  We
find that voids in the distribution of $10^{12}\Msunh$ haloes (expected
galactic magnitudes $\sim M_*$) are almost the same as the voids in
$10^{11}\Msunh$ haloes. Yet, much smaller haloes with masses
$10^{9}\Msunh$ and circular velocities $v_{\rm circ} \sim 20~\kms$
readily fill the voids: there should be almost 1000 of these haloes in a
$20\hMpc$--diameter void. A typical void of diameter $20\hMpc$ contains about 50
haloes with $v_{\rm circ} >50$~km/s.  The haloes are arranged in a
pattern, which looks like a miniature Universe: it has the same
structural elements as the large-scale structure of the galactic
distribution of the Universe. There are filaments and voids; larger
haloes are at the intersections of filaments. The only difference is
that all masses are four orders of magnitude smaller. There is severe
(anti)bias in the distribution of haloes, which depends on halo mass and
on the distance from the centre of the void. Large haloes are more
antibiased and have a tendency to form close to void boundaries.  The
mass function of haloes in voids is different from the ``normal'' mass
function.  It is much steeper for high masses resulting in very few
M33-type galaxies ($v_{\rm circ}\approx 100 \ \kms$). We present an
analytical approximation for the mass function of haloes in voids.

\end{abstract}


\section{Introduction}

Already more than two decades ago it became clear that large regions of
the Universe are not occupied by bright galaxies
\citep{Gregory1978,Joeveer1978, Kirshner81}. Large regions of size $\sim
(10-20)\Mpc$ devoid of galaxies can be clearly seen in all present deep
redshift surveys. The observational discovery was soon followed by the
theoretical understanding that voids constitute a natural outcome of
structure formation {\it via} gravitational instability
\citep{Peebles1982,Hoffman1982}.  Together with clusters, filaments,
and superclusters, giant voids constitute the large-scale structure of
the Universe. In spite of the fact that the voids are important for
understanding the observed structure of the galactic distribution, they
attract much less attention as compared with other elements of the
large scale structure such as clusters of galaxies.

\cite{Einasto1989} were the first to estimate sizes of voids in
different samples of galaxies with measured redshifts.  Voids in the
CfA redshift catalogs were studied by \citet{Vogeley1994}.
\citet{Ghigna1996} made estimates of the void probability function
(VPF) for the Perseus-Pisces region. VPF was estimated for the Las
Campanas redshift survey by
\citet{MuellerLCRS}. \citet{ElAd1997,ElAd2000} studied voids in the
Optical Redshift Survey (ORS) and in the IRAS 1.2-Jy survey. They found that
large voids with radius $\sim 20\Mpch$ occupy about 50\% of the volume
of the Universe. Void distribution in the PSCz catalog was studied by
\citet{Plionis2002} and by \citet{HoyleVogeley} with approximately the same
conclusions regarding the sizes of voids and the fraction of occupied
volume.  For more detailed review of observational efforts see
\citet{Peebles2001}. It should be noted that in spite of significant
efforts, there remain some crucial unresolved issues regarding the
properties of voids. Voids are defined using bright high surface
brightness galaxies. This is quite understandable: finding and
measuring redshifts for dwarf or low surface brightness galaxies
is difficult. For example, the absolute magnitude limit for the sample
used by \citet{Vogeley1994} was $M_B=-19.3$ (assuming the Hubble
constant $h=0.7$). The sample used by \citet{MuellerLCRS} was limited
by $M_R=-20$. In other words, most of the samples are probing voids
using galaxies comparable with the Milky Way.

There were several attempts to find dwarf galaxies in few individual
voids \citep{Lindner1996,Popescu1997,Kuhn1997, Grogin1999}. The overall
conclusion is that faint galaxies do not show a strong tendency to
fill up voids defined by bright galaxies. The limits on absolute
magnitudes of observed galaxies are better than for the large samples,
but not overwhelmingly so. For example, one of the voids studied by
\citet{Kuhn1997} was at a distance of $\sim 3000 \ \kms$, but at that
distance the observational sample was complete only up to $M_B=-18.0$.
For other voids the limit was only $M_B=-20.0$. The strongest
arguments that voids are not populated by dwarf galaxies were given by
\citet{Peebles2001} who points out that the dwarf galaxies in the ORS
catalog follow remarkably close the distribution of bright galaxies:
there are no indications that they fill voids in the distribution of
bright galaxies.  In this case the ORS catalog can ``see'' galaxies
with absolute magnitudes $M_B=-15.5$ up to 10~Mpc distance and within
this distance the voids are clearly empty. A potential problem with
this
statement is that we do not know whether the ORS catalog is missing or
not low luminosity and low surface brightness galaxies. At these
magnitudes the galaxies are likely to have low surface brightnesses.

To summarize, observations indicate that large voids found in the
distribution of bright $(\sim M_*)$ galaxies are empty of galaxies,
which are two magnitudes below $M_*$. The situation at lower limits is
not clear.

Void phenomenon was a target of many theoretical studies
\citep{Einasto1991,Sahni1994, Ghigna1994, Ghigna1996, Friedmann2001,
Arbabi2002,Mathis2002, Bensonetal2003, Antonuccio2002}. VPF was studied
by \citet{Einasto1991} and by \citet{Ghigna1994}.  Our main interest is
not the statistics or the shapes of the voids. We focus on the issue of
emptiness of large voids. Are voids empty or are they filled with dark
matter haloes? What is the structure of the dark matter distribution in
voids?  Does it present a problem for the standard cosmological model?
These are the questions we are trying to address in this
paper. Emptiness of voids is of additional interest in view of
problems of the hierarchical models on small scales: the large
abundance of dark matter satellites of Milky Way size haloes
\citep{KlypinSat1999, Moore1999} and problems with explaining
rotation curves in central parts of dwarf and low surface brightness
(LSB) galaxies \citep[e.g.][]{Moore94, FloresPrimack94,
deBlok2001,BoschSwaters01}. Both problems are on scales of
$10^9-10^{10}\Msunh$, which can be probed by the abundance of dwarf
galaxies in voids.

To some degree the interest for dwarfs in voids is inspired by
\citet{Peebles2001}, who claims that the CDM models have severe
problems: they predict too many dwarfs.  While Peebles did not try to
make any quantitative estimates of the number of galaxies or dark
matter haloes inside voids of large size, the reasoning seems to be
simple and straightforward. At large redshifts the density in a
region, which later will become a void, is not much different from the
average density of the Universe at that redshift. Thus, the
fluctuations grow and haloes collapse. At later times the density in
the region declines and the fluctuations effectively stop growing. The
number of collapsed haloes is preserved in the comoving volume of the
void. This leads to a large number of expected haloes and galaxies in
the void. For example, a void with a density of 1/10 of the average
density of the Universe is expected to have the number density of
galaxies roughly one tenth the average density of galaxies in the
Universe, which gives many galaxies because of the large void
volume. \citet{Mathis2002} argue that gravity removes haloes from
voids. That would reduce the number of haloes and galaxies. Our results
show that this  does not happen and thus cannot solve the
problem.  The simple argument of stopping the growth of fluctuations
and of subsequent dilution of the halo density by void expansion must
work at some scales. In that respect \citet{Peebles2001} is right. The
only question is what is the scale and what happens on larger scales.

Significant progress in understanding the void structure was made
recently by \citet{Mathis2002} and \citet{Bensonetal2003}, who used a
combination of $N$-body simulations with semi-analytical methods to
predict abundance of galaxies in voids in cosmological models. It was
found that the voids are empty even of dwarf
``galaxies''. Unfortunately, the mass resolution in simulations used by
\citet{Mathis2002} and \citet{Bensonetal2003} is low if one wants to
address the issue of dwarfs. The best simulations used by
\citet{Bensonetal2003} had the particle mass $1.4\times
10^{10}\Msunh$, which leads to the minimum halo mass of few times
$10^{11}\Msunh$, which is not much smaller than the mass of our Milky
Way galaxy.  \citet{Mathis2002} had better mass resolution of $3.6\times
10^{9}\Msunh$ giving the minimum halo mass $3.6\times 10^{10}\Msunh$.
One of the goals of our paper is to extend the limit to much smaller
masses to find what happens with real dwarfs in large voids. Indeed,
our mass resolution is almost a hundred times better: we are able to
detect haloes with mass $10^{9}\Msunh$. We also develop analytical
estimates of the mass function of haloes, which we test against
simulations and then apply to much smaller masses.

One significant advantage of the approach used by \citet{Mathis2002} and
\citet{Bensonetal2003} is that they were able to estimate luminosities
of galaxies hosted by dark matter haloes. We do not try to estimate
luminosities. Instead, our high resolution simulations provide the
maximum circular velocities of haloes, which gives us a good idea of what
kind of galaxies we are dealing with.  We note that in any case the
estimates of luminosities in semi-analytical models are still very
uncertain for dwarf haloes: physics of these galaxies is still
poorly understood. This problem is worsened by low mass resolution in
the simulations of \citet{Mathis2002} and \citet{Bensonetal2003}, who
used haloes with as few as 10 particles to track the history of smallest
galaxies.  The estimates of the maximum circular velocities are better
because they do not depend on what is assumed about the star formation
in dwarf galaxies. Yet accurate estimates of circular velocities
require high resolution $N$-body simulations: internal structure of
haloes should be resolved.

In order to get a rough estimate of what luminosities may be expected for
galaxies hosted by haloes in our simulations we present examples of
dwarf galaxies in the Local Group, which have measured circular
velocities and luminosities.  If haloes in simulations host the same
type of galaxies, we should expect the same luminosities.  NGC 6822
and NGC 3109 are irregulars with circular velocities about $60 \kms$ 
and absolute magnitudes $M_B=-15.8$ in the case of NGC 6822
\citep{Hodge1991, Weldrake2003} and $M_B=-15.2$ for NGC 3109
\citep{Mateo1998}. For haloes with the maximum circular velocity of
$60 \ \kms$ the virial mass is about $(2.5-4)\times 10^{10}\Msunh$ if
we assume halo concentrations in the range $10-20$. For galaxies with
this virial mass \citet{Mathis2002} give slightly larger luminosity of
$M_B=-16.3$. Haloes with virial mass $(2-3)\times 10^{11}\Msunh$ play
important role for voids. We find that voids start to fill up with
haloes with masses smaller than this mass.  \citet{Mathis2002} give
absolute magnitudes $M_B$ between $-17.5$ and $-18.5$ for galaxies with
these virial masses. The maximum circular velocities for haloes of this
mass are about $100 \ \kms$, which is comparable with the circular
velocity of the spiral galaxy M33 ($M_B=-18.5$), $v_{\rm circ}=120 \
\kms$ \citep{Corbelli2000}.

We investigate the formation of voids in the standard cosmological
model: a spatially flat $\Lambda$-dominated Universe with
scale-invariant adiabatic Gaussian fluctuations. We use the following
cosmological parameters: $\Omega_{\rm M}=0.3$, $\Omega_{\Lambda}=0.7$,
$\sigma_8=0.9$, and $h=0.7$. The age of the Universe in this model is
approximately $13.5$ Gyrs. These parameters are favored by recent
cosmological observations (e.g.  \citealt{Freedman2001};
\citealt{Riess2001}; \citealt{SpergelMAP}). The normalization
$\sigma_8=0.9$ of our simulations is a rather conservative value
\citep{Bunn1997,Viana1996}. Some recent observational results suggest a
substantially lower normalization or a lower density parameter
$\Omega_{\rm M}$ \citep{Reiprich2002,Viana2002,BahcallDong2003}.
\cite{Pierpaoli2003} found $\sigma_8=0.8$, their Table 1 contains a
compilation of recent estimates of $\sigma_8$.

The plan of the paper is as follows. In Section \ref{sec:simu} we
present our numerical simulations. We discuss briefly the void finding
algorithm which we used to find voids in the distribution of dark
matter haloes of a low resolution simulation and the resimulation of
the regions of selected voids with a higher mass resolution.
Section \ref{sec:analyt} is devoted to the formalism of analytical
predictions of the mass function in voids. In Section \ref{sec:voids}
we discuss the mass and halo distribution in voids.  In Section
\ref{sec:massfunc} we compare the analytical predictions of the mass
function with the mass function measured in the high resolution
simulations of voids.  We discuss how the mass function depends on the
normalization and a changing slope of the power spectrum as recently
proposed by the WMAP collaboration \citep{SpergelMAP}. Finally, in
Section \ref{sec:conclusions} we summarize our results.

\section{Numerical models}
\label{sec:simu}

\subsection{The code}

The Adaptive Refinement Tree (ART) $N$-body code of \cite{Kravtsov1997}
was used to run all numerical simulations analyzed in this paper. This
code uses Adaptive Mesh Refinement technique to achieve high resolution
in the regions of interest. The computational box is covered with a
uniform grid which defines the lowest (zeroth) level of resolution.
The code then reaches high force resolution by recursively refining all
high density regions using an automated refinement algorithm.  The
shape of the refinement mesh can thus effectively match the geometry of
the region of interest. This algorithm is well suited for simulations
of a selected region in a large computational box, as in the
simulations presented below. During the integration, spatial refinement
is accompanied by temporal refinement.  Namely, each level of
refinement, $l$, is integrated with its own time step $\Delta
a_l=\Delta a_0/2^l$, where $\Delta a_0$ is the global time step of the
zeroth refinement level.  In addition to spatial and temporal
refinement, simulations described below also use a non-adaptive mass
refinement algorithm to increase the mass (and correspondingly the force)
resolution inside a specific region \citep{Klypin2001}.

We start with running a low resolution simulation with $128^3$
particles covering the whole computational box. For that we make a
realization of the initial spectrum of perturbations with $1024^3$
particles in the simulation box. Initial coordinates and velocities of
the particles are then calculated using all waves ranging from the
fundamental mode $k=2\pi/L$ to the Nyquist frequency $k=2\pi/L\times
N^{1/3}/2$, where $L$ is the box size and $N=1024^3$ is the number of
particles in the simulation. Then we replace every $8^3=512$ particles
with particles of larger mass. A large-mass (merged) particle is
assigned a velocity and displacement equal to the average velocity and
displacement of the small-mass particles.  Once a simulation with
$128^3$ particles is completed, we select voids and identify all
particles inside the voids.

Then we restart the simulation keeping high mass resolution for
particles in voids.  Particles outside the high resolution region are
merged in several steps so that the high resolution region
corresponding to $1024^3$ particles is surrounded by shells with
resolution corresponding to $512^3$ and $256^3$ particles. The
remaining part of the simulation box is simulated in low mass
resolution corresponding to $128^3$ particles.  High force resolution
is only achieved inside the region with high mass resolution. The
details of this multi-mass technique are described by
\cite{Klypin2001}.

\subsection{Simulations}

In order to study the formation of large voids, the simulation box
should be sufficiently large; we use a cube of 80\hMpc\ on a side.
This is sufficient because we are not interested in a statistics of
large voids, which would require a significantly larger volume.  Our
main interest is in the structure of a typical $\sim 10\Mpch$--radius 
void. The 80\hMpc\ box is large enough for that. We focus on the
formation of small structural elements (haloes and filaments) inside
voids, for which we need the highest possible mass resolution.

The limitation of $1024^3$ particles gives the $4.0 \times 10^7\Msunh$
per particle.  This leads to the minimum halo mass of $10^9\hMsun$. A
halo with this mass has the maximum circular velocity $\sim 20~\kms$.
Inside the high resolution region we reach the force resolution (one
cell) of $0.6 \hkpc$. We make 250000 time-steps on the highest
resolution level.

The identification of haloes is always a challenge. The widely used
halo-finding algorithms, the friends-of-friends (FOF) and the
spherical overdensity, both discard ``haloes inside haloes'', i.e.
satellite haloes located within the virial radius of larger haloes. We
have developed two algorithms that do not suffer from this drawback: the
hierarchical friends-of-friends (HFOF) and the bound density maxima
algorithms (BDM, see \citealt{Klypin1999}).

The algorithms are complementary. They find essentially the same
haloes. Thus we believe that the algorithms are stable and capable of
identifying all dark matter haloes in our simulations.  The advantage
of the HFOF algorithm is that it can handle haloes of arbitrary shape,
not just spherical haloes.  The advantage of the BDM algorithm is that
it describes the physical properties of the haloes better by
identifying and removing unbound particles.  In particular it estimates
not only the mass of a halo, but also its maximum ``circular velocity'',
$v_{\rm circ}=\sqrt{GM/R}$.  This is the quantity which is more
meaningful observationally.  Numerically, $v_{\rm circ}$ can be
measured more easily and more accurately than the mass. In order to
compare the velocity function of haloes measured in simulations with
analytical predictions one has to convert the virial masses into
circular velocities assuming an NFW density profile \citep{Got1999Sesto}.

\subsection{Finding voids}

In order to identify voids, we start with construction of the minimal
spanning tree for selected haloes. Typically we select haloes with mass
larger than $2\times 10^{11}\Msunh$, but different criteria were also
used.  Then we search on a grid with mesh size $0.6 \hMpc$ for the point
in the simulation box which has the largest distance $R_1$ to the set
of haloes. This is the centre of the largest void the radius of which
is $R_1$. We exclude this void and search again for a point with the
largest distance to the set. This gives the second largest void and so
on. The algorithm is similar to that used by
\cite{Einasto1989}. \cite{ElAd1997} use a somewhat more complicated
search algorithm based on ``wall'' and ``field'' galaxies, where field
galaxies are allowed to be also in voids.

In principle, our algorithm (as the algorithm of
\citealt{ElAd1997}) allows for the construction of voids with arbitrary
shape: the starting point is a spherical void which can be extended by
spheres of lower radius which grow from the surface of the void into
all possible directions. However, in the following analysis we have
restricted ourselves to spherical voids to avoid ambiguities of
the definition  of allowed deviations from spherical shape.

\section{Analytical approximations for the mass function in a low
             density region}
\label{sec:analyt}

In this section we provide a formalism for predicting the mass
function of dark matter haloes in voids. There are different ways of
constructing the mass function in voids.  Using the barrier-crossing
formalism of \citet{BondCole1991} and \citet{LaceyCole1993},
\citet{MoWhite1996} generalized the Press-Schechter approximation so
that it can be applied to over- and under-dense regions.
The validity of this approach has been verified against $N$-body
simulations by \citet{LemsonKauffmann1999}. A better
approximation is provided by \citet[][ST]{Sheth2002}. We compare
predictions based on this approximation with our $N$-body results and
find that it does not provide an accurate fit to the results of
simulations. This motivates us to develop our own approximation.

For completeness, we start with presenting the constrained ST
approximation.  If \rhovoid~ is the mean density of matter in a void
and $\sigma(R)$ is the rms density fluctuation at comoving scale $R$,
then the number density $n_{\rm c,ST}$ of haloes with mass
$M(R)=4\pi\rho_{\rm b} R^3/3$ is given by

\begin{eqnarray}
    n_{\rm c, ST}(M) &=& - \left( \frac{2}{\pi} \right)^{1/2}
    \frac{\varrho_{\rm void}}{M}
    \frac{| T(\sigma^2 | \sigma_0^2)|}{(\sigma^2 -\sigma_0^2)^{3/2}}
    \frac{\sigma {\rm d} \sigma}{{\rm d} M} \nonumber \\
    & \times & \exp \left[ - \frac{[B(\sigma^2) - B(\sigma_0^2)]^2}{2
    (\sigma^2-\sigma_0^2)} \right]
    \label{eq:cST}
\end{eqnarray}
where
\begin{equation}   \label{v15}
    T(\sigma^2 | \sigma_0^2) = \sum_{n=0}^5 \frac{(\sigma_0^2 - \sigma^2)^n}{n!}
    \frac{\partial^n [B(\sigma^2) - B(\sigma_0^2)]}{\partial (\sigma^2)^n}
\end{equation}
(note the correction of the typo with respect to \citealt{Sheth2002};
R. Sheth, private communication). Here
\begin{eqnarray}
    B(\sigma^2) &=& a^{1/2} \delta_{\rm c}[1+ \beta (a \delta_{\rm c}^2/\sigma^2)^{-\alpha}] ,
    \label{v16}  \\
    B(\sigma_0^2) &=& a^{1/2} \delta_0 [1+ \beta (a \delta_0^2/\sigma_0^2)^{-\alpha}].
    \label{v17}
\end{eqnarray}
The parameters $\alpha=0.615$, $\beta=0.485$ are given by the
ellipsoidal dynamics, while $a=0.707$ is adjusted by comparison with
simulations.
The parameter $\delta_0$ is the linear underdensity of the void
corresponding to the actual nonlinear underdensity at the moment at
which we measure the mass function ($z=0$ in our case). It is calculated
from the spherical top-hat model \citep{Sheth2002}:

\begin{eqnarray}
    \delta_0 (\delta) &=& \frac{\delta_{\rm c}}{1.68647} \left[ 1.68647
    - \frac{1.35}{(1+\delta)^{2/3}} \right.
    \nonumber \\
    &-& \frac{1.12431}{(1+\delta)^{1/2}}
    + \left. \frac{0.78785}{(1+\delta)^{0.58661}} \right].     \label{v13}
\end{eqnarray}
Here $\delta$ denotes the mean density contrast in the
void, $\delta=\varrho_{\rm void}/\varrho_{\rm b} - 1 = \Delta (R_{\rm
void})/\Omega_{\rm M} - 1$, where $\Delta = 3 M(R) /(4 \pi
R^3 \varrho_{\rm c})$ is the mean density in a sphere of radius $R$.
We will assume that the density in a void does not depend on the
distance from the centre of the void. Later we will see that this not
exactly true, but it is a reasonable starting point.
In the expression above, $\varrho_{\rm b}$ is the background density,
$\delta_{\rm c}$ is the characteristic density for collapse as
predicted by linear theory according to the spherical collapse model
(for our cosmology, $\delta_{\rm c}=1.676$, see \citealt{Lokas2001}).

The parameter $\sigma_0$ is defined as the linear rms fluctuation on
the scale of the void. It is estimated using the linear power spectrum
and the top-hat filter with radius $R_0$ defined by
$R_0^3=(1+\delta)R^3_{\rm void}$, where $R_{\rm void}$ and $\delta$
are the radius and the mean density contrast of the void. The
parameter $\sigma_0$ is essential for this approach. It provides
truncation for objects with large mass: as the amplitude of
perturbation $\sigma$ approaches the amplitude of perturbation of the
void, the number density of haloes goes to zero. The effect of
truncation is ``felt'' even for much smaller objects.

The formalism of the constrained mass function in the version proposed
by \cite{Sheth2002} is rather complicated and arbitrary in a sense that the
series in equation (\ref{v15}) is not well motivated and the parameters of the
model are adjusted by comparison with $N$-body simulations.  We introduce
an alternative and in our opinion much more straightforward and
natural method of predicting the mass function in voids. As we will
show in Section \ref{sec:massfunc}, it also reproduces our simulated
mass functions more accurately. Our method is based on the assumption
that the evolution of matter distribution in a void proceeds
effectively as it would in a Universe with cosmological parameters
similar to those of the void. We treat the void as a Universe with a
density parameter $\Omega_{\rm M,void} = \Delta(R_{\rm void})$ with
$\Delta(R_{\rm void})$ measured from the simulations.

The growth of perturbations in such a Universe is also slower than in
the whole Universe. In order to take this into account we change the
normalization of the power spectrum in the void by assuming a new
value of $\sigma_{8, {\rm void}}$, related to the background
$\sigma_8$ by
\begin{equation}   \label{v6}
   \sigma_{8, {\rm void}} = \sigma_8 \frac{D(a_{\rm i})}{D(a=1)} \frac{D_{\rm
   void} (a=1)}{D_{\rm void} (a_{\rm i})}
\end{equation}
where $D(a)$ and $D_{\rm void}(a)$ are the linear growth factors of density perturbations
in the background and in the void respectively, normalized so that for $\Omega_{\rm M} =1$
and $\Omega_\Lambda=0$ we
have $D(a)=a$.  We assume that at some initial $a_{\rm i} = 1/(1+z_{\rm
i})$ $(z_{\rm i} \approx 1000)$ the rms fluctuations in the background
and in the void were equal.

For a flat model with $\Omega_{\rm M} + \Omega_\Lambda=1$ describing our background Universe
the growth factor is given (\citealt{Silveira1994}, corrected for typos) by
\begin{equation}    \label{v7}
    D_{\rm flat}(a) = a \ \ _2 F_1 \left[ \frac{1}{3},
    1, \frac{11}{6},  a^3
    \frac{\Omega_{\rm M}  -1}{\Omega_{\rm M} } \right]
\end{equation}
where $_2 F_1 $ is a hypergeometric function. For non-flat cosmologies,
such as those of the voids, $D(a)$ has to be calculated numerically using
\citep{Heath1977,Carroll1992}
  $  D(a) =[5 \Omega_{\rm M}/(2 a f(a))] \int_{0}^{a} f^3 (a) {\rm d} a $,
where
  $  f(a) = \left[ 1+ \Omega_{\rm M} \left(\frac{1}{a}
    -1\right) + \Omega_\Lambda \left(a^2 - 1\right) \right]^{-1/2}. $
We then use unconstrained ST approximation \citep{Sheth1999} to find
the mass function of haloes in voids:
\begin{eqnarray}
    n_{\rm ST}(M) &=& - \left( \frac{2}{\pi} \right)^{1/2}
    A \left[ 1 + \left( \frac{a \delta_{\rm c}^2}{\sigma^2} \right)^{-p}
    \right] a^{1/2}  \label{v11} \\
    & \times& \frac{\varrho_{\rm b}}{M}
    \frac{\delta_{\rm c}}{\sigma^2} \frac{{\rm d} \sigma}{{\rm d} M}
    \exp \left( - \frac{a \delta_{\rm c}^2}{2 \sigma^2} \right),   \nonumber
\end{eqnarray}
where $A=0.322$, $p=0.3$ and $a=0.707$. Here we
take $\delta_{\rm c}=1.62$, which is appropriate for the `open Universe' parameters
of the voids in our simulations $\Omega_{\rm M, void} = 0.03-0.05$ and
$\Omega_\Lambda=0.7$ (see \citealt{Lokas2001}).

\section{Voids in  simulations}
\label{sec:voids}

\subsection{Mass distribution in voids}
\begin{figure}
\includegraphics[width=\columnwidth]{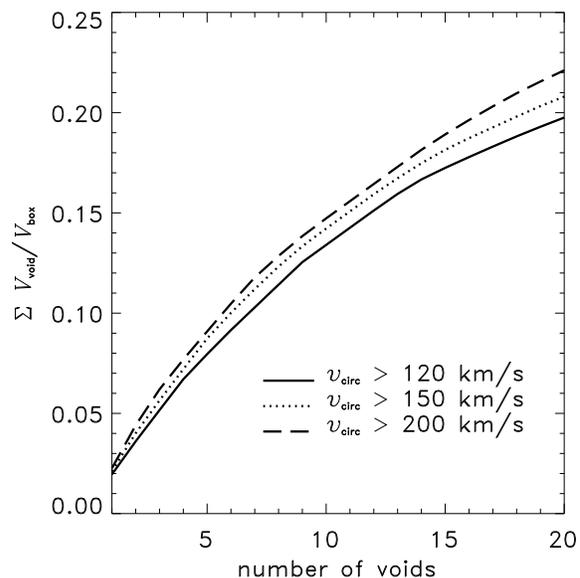}
\caption{Cumulative fraction of the volume occupied by 20 largest
voids in the distribution of haloes with different limiting
circular velocities. More massive haloes define slightly larger voids.}
\label{fig:voidvolumes}
\end{figure}

In the simulation box of size 80 \hMpc\ we found 1348 dark matter
haloes with circular velocities larger than 200 km/s and 2518 with
circular velocities larger than 120 km/s. We search for spherical
non-overlapping voids in the distribution of the dark matter haloes.
  The 20 largest voids have radii
larger than 8.81 \hMpc\ when the limiting circular velocity of haloes
is set to $v_{\rm circ} > 200$ km/s while larger than 8.26 \hMpc\ for
haloes with $v_{\rm circ} > 120$ km/s. In Figure \ref{fig:voidvolumes}
we show the cumulative fraction of volume occupied by the 20 largest
voids. The voids in the distribution of more massive haloes tend
to be larger, but the difference is small as long as the difference in
mass is not very big. A similar behaviour has been found by
{\cite{Arbabi2002}}. This indicates that there are only rare cases
when a given void is divided into two parts if the threshold for the
circular velocity (or mass) of the objects defining the void is
reduced. This statement is supported by our conclusion that there is a
tendency to find more massive haloes in the outer part of voids.

Five voids were resimulated with mass resolution $4.0 \times 10^7
\hMsun$. In the low resolution simulation their radii were $R_{\rm
  void} = 11.6 {\hMpc}$, $10.8 {\hMpc}$, $9.4 {\hMpc}$, $9.1 {\hMpc}$,
$9.1 {\hMpc}$. With the high resolution mentioned above we
resimulated regions with 10\% larger radii, so that the objects which
define the borders of the voids have been also resimulated. The void
finding algorithm assumes point-like objects. However, these objects
have a certain size. Moreover, they could be surrounded by satellites
with smaller masses. Since we do not want to include in our analysis
the objects themselves or their satellites (which in the sense of the
definition do not belong to the void) we have assumed a void radius of
10 \hMpc\ or 8 \hMpc\ for the large and small voids respectively.

\begin{figure}
\vspace{-1cm}
\includegraphics[width=\columnwidth]{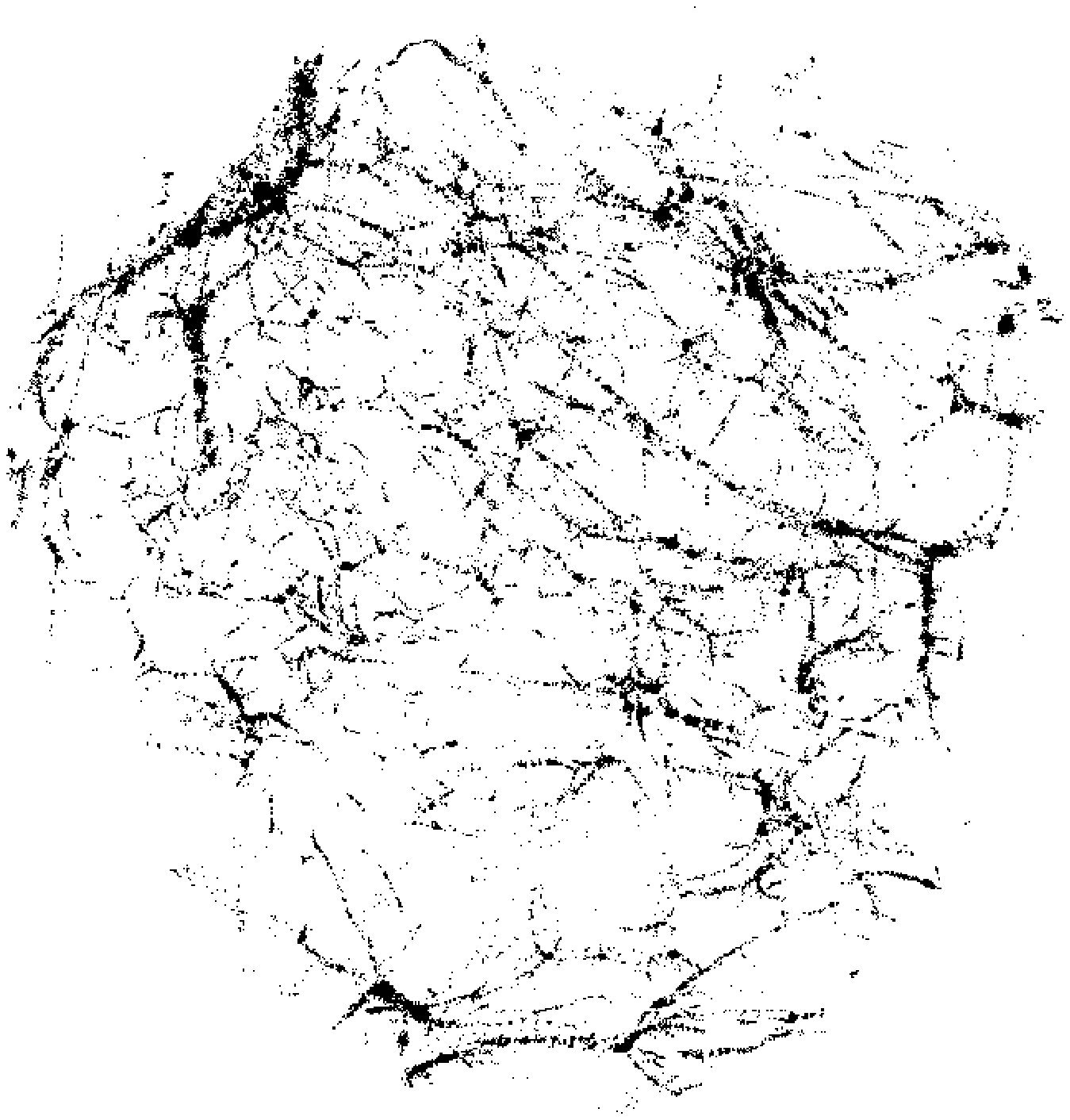}\\[-1.5cm]
\includegraphics[width=\columnwidth]{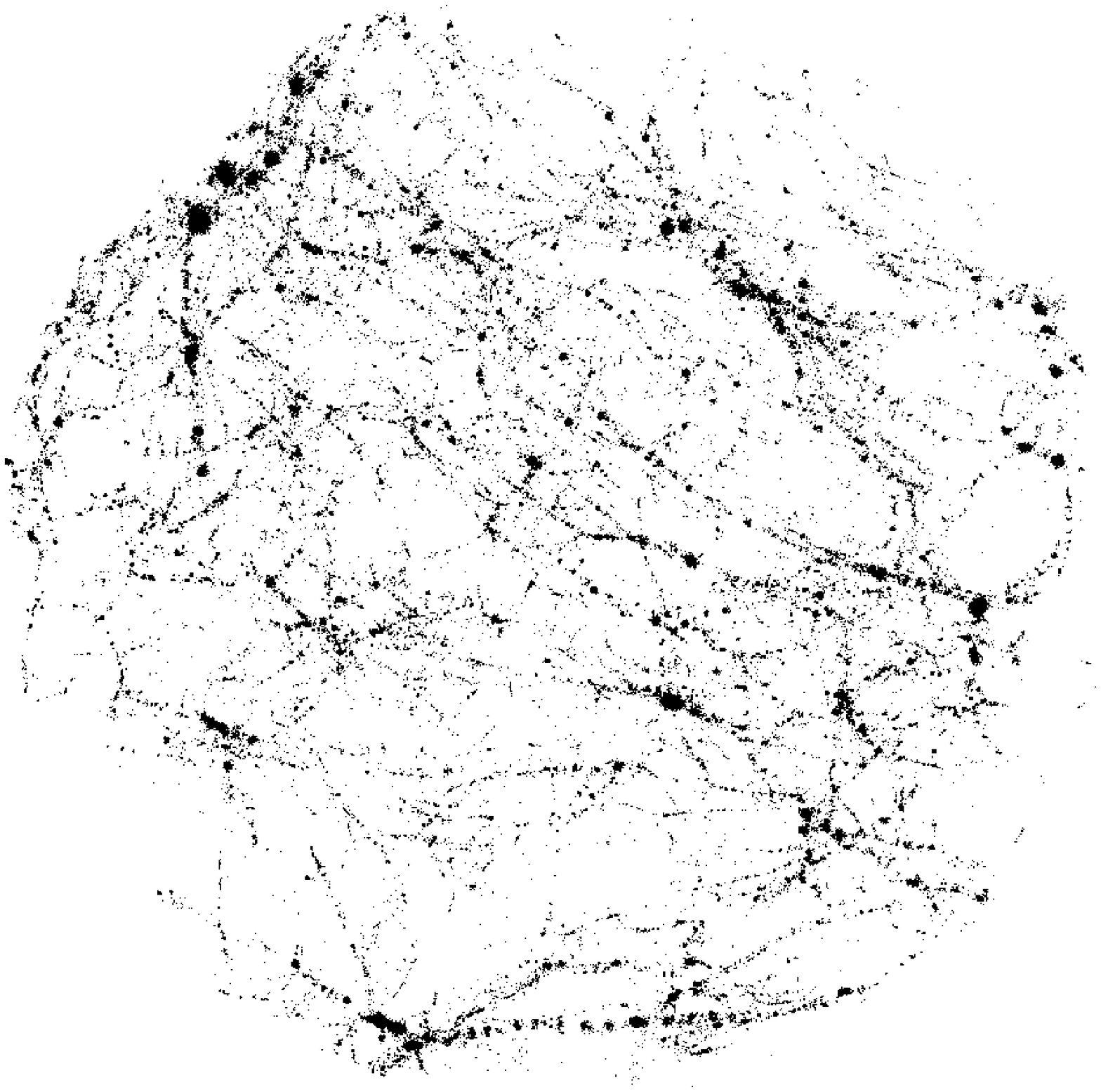}
\vspace{-1cm}
\caption{The distribution of dark matter inside a large void. Bottom
  panel shows the distribution of matter in a void with radius $10
  \hMpc$ at redshift $z=0$. Top: all particles belonging to the void
  are traced back to redshift $z=2$ and are shown in comoving
  coordinates. The distribution looks like normal large-scale structure
  in the distribution of galaxies: the haloes are in filaments, there
  are ``voids'' and most massive objects are at intersections of large
  filaments. Yet, the whole region is a void: there is no Milky Way
  size halo in the picture. Numerous knots along the filaments are
  haloes with masses $10^{9}\Msunh$.  Large haloes have masses of few
  times $10^{10}\Msunh$.}
\label{fig:void}
\end{figure}

In the bottom part of Figure \ref{fig:void} we show a sphere of radius 10
\hMpc\ centered on the void of radius 10.8 {\hMpc}. This void does not
contain any halo of mass greater than $2.0 \times 10^{11}$ {\hMsun}.
The progenitor of this void at redshift $z =2$ is not spherical. It is
much smaller in comoving coordinates, i.e.  the density contrast of the
void with respect to the mean density was much smaller at high
redshifts. Note, however, that this does not prove the statement that
voids become more spherical during evolution, because the selected
voids are spherical by definition at $z=0$. The void and its progenitor
are shown in the same projection in comoving coordinates. The obvious
shift of the centre of the void means that the void not only
expands anisotropically with respect to the background but also that the
whole void moves with respect to the comoving coordinate system. It is
interesting to see that there is already a huge number of dense
filaments at redshift $z = 2$ which should be observable in the
Lyman $\alpha$ forest.

\begin{figure}
\includegraphics[width=\columnwidth]{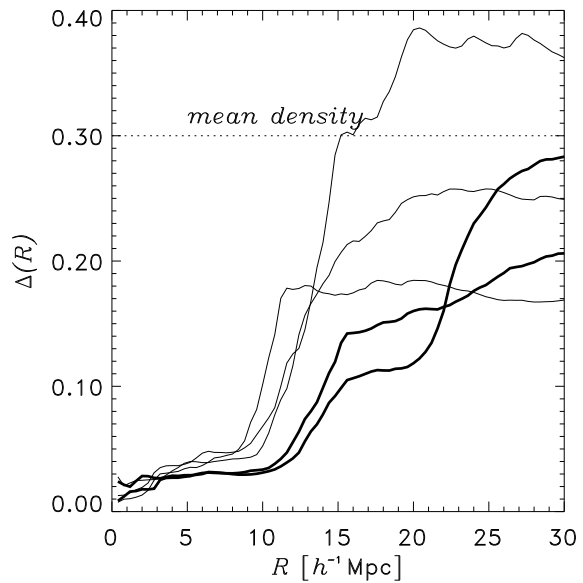}
\caption{
The mean density in spheres of radius $R$ centered on the centre of
each of the five voids. The two thick solid lines denote the voids with
radius $10 \hMpc $, the three thin solid lines denote voids with radius
$8 \hMpc $.
}
\label{fig:density}
\end{figure}

Inside the void at redshift $z=0$ we find almost the same structures as
seen in simulations of large parts of the Universe: empty regions,
filaments and matter concentrations at the points where filaments
join. However, all masses are scaled down by a factor of several orders
of magnitude.  In the crossing points of filaments we find instead of
huge dark matter haloes hosting clusters of galaxies only small haloes
which might host dwarf galaxies. Along the filaments even smaller
haloes are situated.  The three-dimensional distribution of matter
seems to be non-uniform.  Large nodes of filaments seem to be situated
nearer to the border of the void than to its centre. This visual
impression is supported by the distribution of dark matter (cf. Figure
\ref{fig:density}) and haloes (cf. Figure \ref{fig:halo_distrib}).

Let us first consider the mean density in spheres centered at the
void's centre. We express the mean density in units of the critical
density, i.e. in a way similar to the (constant) $\Omega$ parameter:
$\Delta = 3 M(R) /(4 \pi R^3 \varrho_{\rm c})$ where $M(R)$ is the
total mass inside a sphere of radius $R$ centered at the centre of the
void and $\varrho_{\rm c}$ is the critical density. As one can see in
Figure \ref{fig:density}, inside a void the density increases slightly
with radius and is typically a factor of 10 smaller than the mean
density $\Omega_{\rm M} = 0.3$. Our voids have smaller densities than
those described by \cite{Friedmann2001}, who found that voids typically
have half the mean density.

Large differences can be seen in the environment of voids.  Most of the
voids are situated in regions of low density. Up to the radius of 30
\hMpc\ the mean density in the spheres is still well below the mean
density of the Universe.  However, the most prominent structure of the
simulation, a galaxy cluster of mass $2 \times 10^{15} \hMsun$, is
close to void 4. Therefore, the density outside this void rapidly
increases so that the mean density of a sphere of radius 20 \hMpc\
centered on this void is already well above $\Omega_{\rm M} = 0.3$.  On
the contrary, the most prominent clusters in the box are at distances
of 51 \hMpc\ resp. 47 \hMpc\ from the center of void 5. Therefore, the
mean density in spheres centered on the centre of void 5 remains below
0.2 up to radii 30 \hMpc\ and reaches 0.3 only for radii above 50
\hMpc\ .  We did not find any significant differences in the
inner structure of the voids so we conclude that the environment of the
voids has no influence on the void itself.

\subsection{Haloes in voids}
\label{sec:halos}

Visual inspection of the void simulations shows plenty of haloes in the
void. In Section \ref{sec:simu} we have described how haloes can be
identified in dark matter simulations. At first glance we can see that
haloes are not homogeneously distributed in the void (Figure
\ref{fig:void}), large haloes seem to be concentrated in the outer parts
of the void. We want to quantify this statement.

\begin{figure}
\includegraphics[width=\columnwidth]{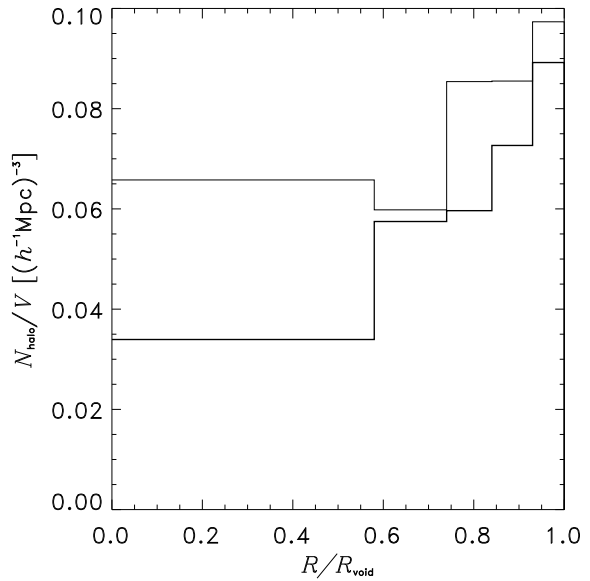}
\caption{Mean number density of haloes with circular velocities 55 km/s $ <
v_{\rm circ} < 120$ km/s (thick line) and 20 km/s $ < v_{\rm circ} < 55$
km/s (thin line) in shells of equal volume ($V_{\rm void}/5$) for five
voids. The radius of the shell is normalized to the void radius.}
\label{fig:halo_distrib}
\end{figure}

We divided our sample of haloes in the five voids into two subsamples
containing 164 haloes with circular velocities 55 km/s $ < v_{\rm circ}
< 120$~km/s and 207 haloes with circular velocities 20 km/s $ < v_{\rm
circ} < 55$~km/s. Each void was divided into five shells of equal
volume. In Figure \ref{fig:halo_distrib} we show the mean number
density of haloes in the five shells the radius of which is normalized
to the void radius. The thick line in the Figure corresponds to the
haloes with higher circular velocities, while the thin one to those
with lower velocities.  One can clearly see that there is a tendency
for the haloes to concentrate more at the outer parts of the
voids. This tendency is more pronounced for the more massive haloes
(thick line), their density in the outer shell is almost a
factor of three higher than in the central sphere.

\begin{figure}
\includegraphics[width=\columnwidth]{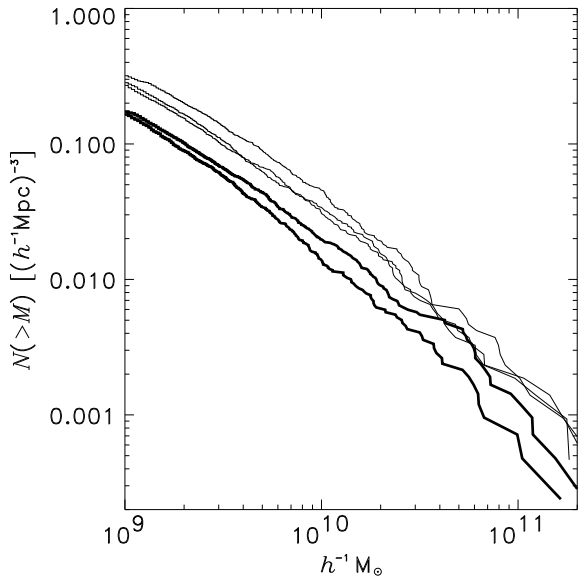}
\caption{Mass functions of haloes in five simulated voids. The two
  thick solid lines correspond to voids with radius $R_{\rm void}=10
  \hMpc $ and mean density $\Delta(R_{\rm void})= 0.03$, while the
  three thin solid lines denote voids with radius $R_{\rm void}=8
  \hMpc $ and mean density $\Delta(R_{\rm void})= 0.04$ (see Figure
  \ref{fig:density}). }
\label{fig:mass_func_simu}
\end{figure}

Down to the limit of $10^{9}\Msunh$ we find thousands of haloes in the
simulation. Now we are interested in the mass function of haloes in
different voids and its dependence on the mean density in the void. We
select all haloes inside the assumed void radius $R_{\rm void} = 10
\hMpc$ or $R_{\rm void} = 8 \hMpc$ depending on the void and estimate the
mass function of haloes in each of the voids.
Figure~\ref{fig:mass_func_simu} shows the five mass functions
measured in the five simulated voids. The mean dark matter density in
the voids (see Figure \ref{fig:density}) is about $\Delta(R_{\rm
void})= 0.03$ for the larger voids with $R_{\rm void} = 10 \hMpc$ and
$\Delta(R_{\rm void})= 0.04$ for the smaller voids with $R_{\rm void} =
8 \hMpc$.

\begin{figure}
\includegraphics[width=\columnwidth]{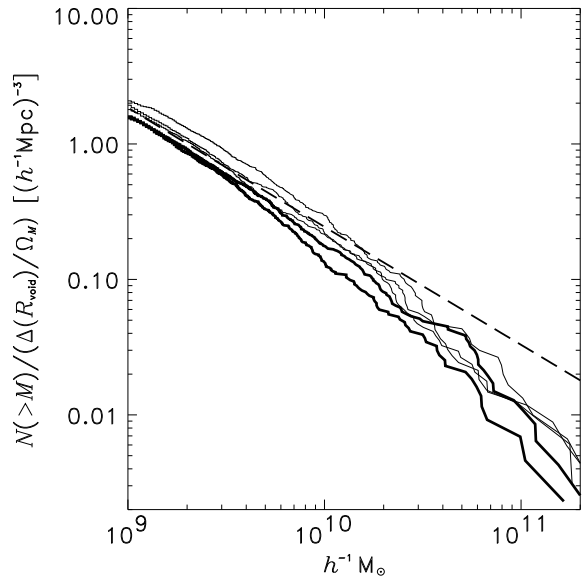}
\caption{The same mass functions as in Figure \ref{fig:mass_func_simu} but
scaled by $\Delta(R_{\rm void})/\Omega_{\rm M}$. The dashed line
denotes the mass function of the field according to the Sheth-Tormen
formalism, thin and thick solid lines correspond to voids
with different radii as in Fig. \ref{fig:mass_func_simu}.  }
\label{fig:mass_func_simu_n}
\end{figure}

The different mean (under)densities of the simulated voids result in
different mass functions. The higher the density in voids the higher is
also the number density of haloes. The number density of haloes in
voids is about an order of magnitude smaller than in the whole box as
expected due to mean density in voids which is also about an order of
magnitude smaller \citep{GottloeberLokas2002}. This can be also seen in
Figure \ref{fig:mass_func_simu_n} where we have shown the mass function
of field haloes obtained by the Sheth-Tormen formalism. This mass
function is in excellent agreement with the mass function measured in
the whole 80 \hMpc\ box \citep{GottloeberLokas2002}. In Figure
\ref{fig:mass_func_simu_n} we have scaled the mass functions in voids
with the mean density contrast measured in the void with respect to the
mean density of matter in the universe, $\Delta(R_{\rm
void})/\Omega_{\rm M}$. Now the scatter between the mass functions of
the different voids is smaller. Note, that the shape of the mass
function in voids is steeper than in the whole box.  One can clearly
see that after rescaling the overall mass function with the density in
voids, the less massive halos are as abundant in voids as the general
mass function predicts. The more massive halos, on the other hand, are
deficient in the voids.

\section{The mass function in voids}
\label{sec:massfunc}
\subsection{Comparison of simulated and analytical mass functions}
\label{sec:compare}

\begin{figure}
\includegraphics[width=\columnwidth]{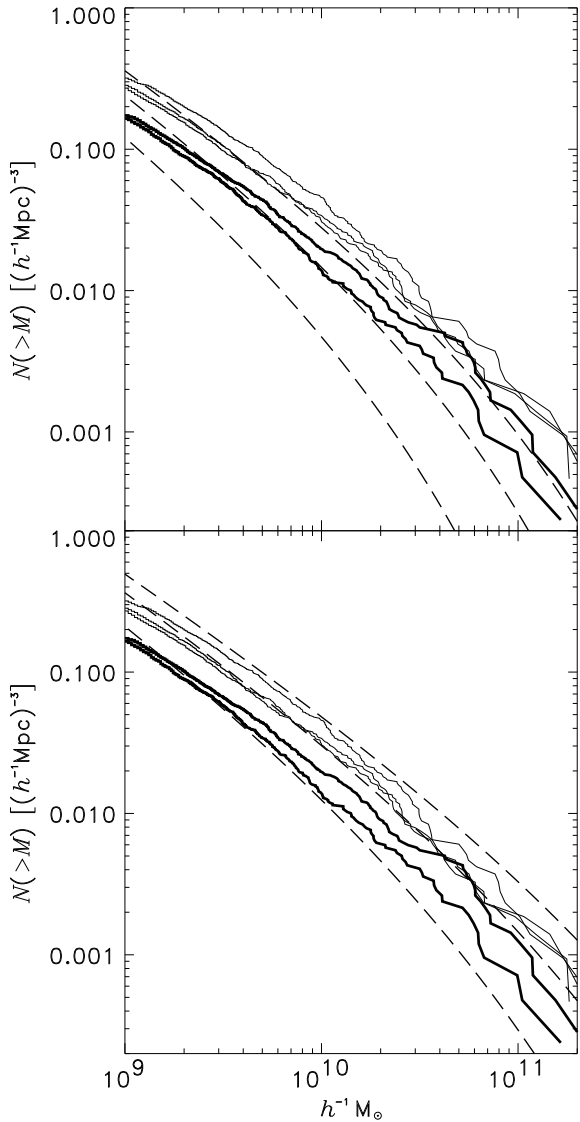}
\caption{Comparison of the mass functions in five simulated voids
  (thin and thick solid lines correspond  to voids with
  different radii as in Fig. \ref{fig:mass_func_simu}) to the
  predictions (dashed lines) based on constrained ST mass function (top
  panel) and rescaling the unconstrained ST mass function (bottom
  panel).  The dashed curves from the highest to the lowest are for
  $\Delta(R_{\rm void})=\Omega_{\rm M,void} = 0.05, 0.04, 0.03$.}
\label{fig:mass_func_comp}
\end{figure}

The top panel of Figure~\ref{fig:mass_func_comp} shows predictions
obtained using constrained Sheth-Tormen mass function
equations (\ref{eq:cST})-(\ref{v17}) with $\Delta (R_{\rm void}) = 0.05,
0.04, 0.03$ and assuming the size of the void $R_{\rm void}=10 h^{-1}$
Mpc (smaller sizes of voids result in steeper mass functions).  The
analytical results presented here differ from those of
\cite{GottloeberLokas2002} in that they assumed $a=0.5$ in equations
(\ref{v16})-(\ref{v17}) to fit the simulated mass function better
while here we keep the value $a=0.707$ advertised by Sheth and Tormen
everywhere. The constrained mass functions do not provide good fits to
the $N$-body results. They are too steep and are a factor of 2--10 below
the simulations. They are especially bad for voids with very low densities.

In the bottom panel in Figure~\ref{fig:mass_func_comp} we compare the
simulated mass functions with the predictions based on rescaled
mass function given by equation~(\ref{v11}). To make the analytical
predictions we use the same values of the mean density as in the
simulated voids: $\Omega_{\rm M,void} = \Delta (R_{\rm void}) = 0.05,
0.04, 0.03$. For voids with these densities we get the corrected values
for the normalization of the power spectrum $\sigma_{8,{\rm void}}=
0.25, 0.20, 0.14$ respectively. It is clear that the rescaled mass
functions provide much better fits to the simulations. Yet, the fits
are not very accurate.  One may try to improve the fits by changing
parameters $A$, $p$ and $a$ in equation~(\ref{v11}). After all, values of
these parameters to some extent were tuned to fit $N$-body
simulations.  However, this is not a good way to
improve the approximation:  predictions are quite stable. For example,
the values of $A$, $p$ and $a$  suggested by
\cite{Jenkins2001} do not change our results significantly and do
not produce better agreement with simulations.

\begin{figure}
\includegraphics[width=\columnwidth]{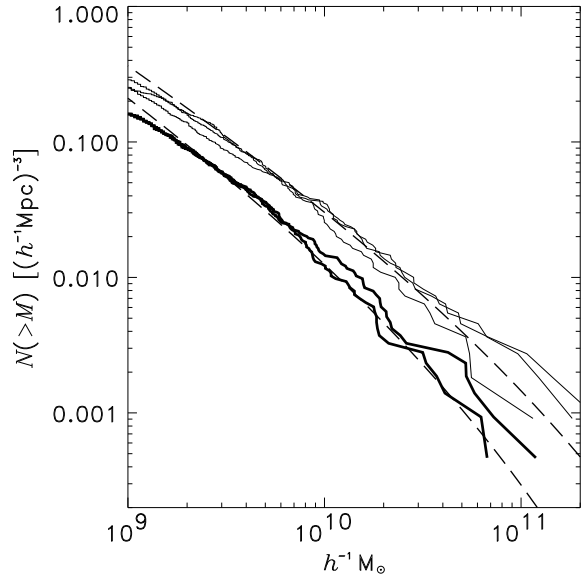}
\caption{The solid curves are mass functions in simulations of
  voids. Void radii are reduced by 20\% as compared with the void radii
  in Fig. \ref{fig:mass_func_simu}. The two thick solid lines
  correspond here to the radius $R_{\rm void}=8 \hMpc $, while the three
  thin solid lines correspond to the radius $R_{\rm void}=6.4 \hMpc
  $. Dashed curves show predictions from the rescaled ST mass
  functions for $\Omega_{\rm M,void} =0.03$ and $0.04$. The analytical
  predictions and numerical results are now in a very good agreement.}
\label{fig:mass_func_volume}
\end{figure}

The low accuracy of the approximation can be traced to the fact that
we assume a constant density of the dark matter inside a void. At the
same time simulated voids have density, which visibly increases close
to the void boundaries. The number density of haloes (especially
massive ones) is very sensitive to the average dark matter density as
clearly illustrated by Figure~\ref{fig:halo_distrib}. This
inconsistency in treatment of the voids is the main reason for poor
quality of the fits. The tendency of more massive haloes to
concentrate in the outer part of the voids is also in agreement with
the slowly increasing cumulative volume fraction of voids defined by
samples of haloes with different minimum circular velocities
(cf. Figure~\ref{fig:voidvolumes}). In fact, if those haloes were
uniformly distributed one would more often expect that one of the big
voids will be divided into two smaller ones if one decreases the
threshold of the halo mass or circular velocity defining the voids.
This is not the case as one can see from the small differences in the
volumes of the largest voids defined in the set of haloes with
different circular velocities (see Figure~\ref{fig:voidvolumes}).

In order to reduce the effect of the varying density, we find the
simulated halo mass function using void radii which are 20\% smaller
than their actual radii.  Figure \ref{fig:mass_func_volume} shows the
results. The full curves represent the mass functions measured in
the five voids with the reduced radii. They are now steeper at the high
mass end. The dashed curves show the predictions based on the
rescaled Sheth-Tormen approximation (the same as in the bottom panel of
Figure~\ref{fig:mass_func_comp}). The agreement between the analytical
predictions and numerical results is now significantly improved.

\subsection{Dependence of the halo mass function in voids on different parameters}
\label{sec:predict}

 Even now the normalization of the power spectrum of density
perturbations has some uncertainties. Observational constraints before
recent WMAP results give uncertainty of about 10\% for the parameter
$\sigma_8$ (see \citealt{BahcallDong2003} for SDSS estimates and for a
discussion of other measurements which have a tendency to predict low
values of $\sigma_8$).  However, the WMAP results \citep{BennettMAP}
again favour $\sigma_8 = 0.9$ if the perturbation spectrum is scale
invariant with $n=1$.  The mass function in voids is affected by all
these uncertainties: it becomes steeper with decreasing normalizations
of the background power spectrum (with other cosmological parameters
unchanged).

\begin{figure}
\includegraphics[width=\columnwidth]{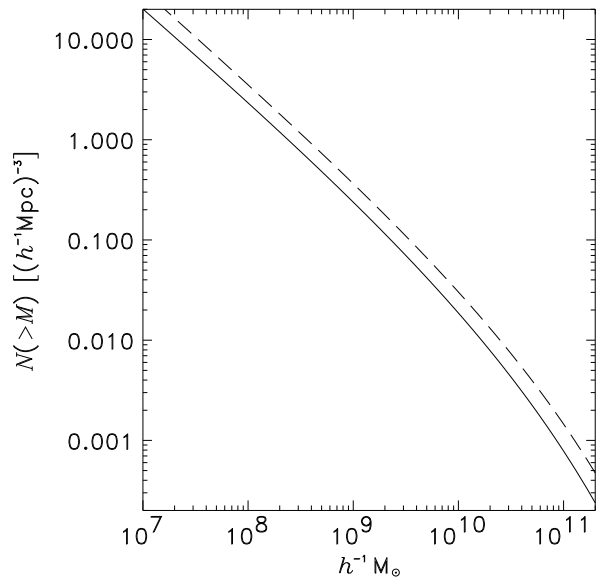}
\caption{Rescaled ST mass function for $\Omega_{\rm M,void}
= 0.04$ and $\sigma_8=0.9$ (dashed line) in comparison to the mass
function obtained for $\sigma_8=0.84$ and scale-dependent spectral
index $n=0.93-0.031 \ln (k/k_0)$ (solid line) as proposed by WMAP
\citep{SpergelMAP}.}
\label{fig:normal}
\end{figure}

The combination of WMAP results with other CMB measurements, 2dF and
Lyman $\alpha$ forest results indicate a lower normalization of
$\sigma_8=0.84$ together with a scale-dependent slope of the primordial
power spectrum $n=0.93-0.031 \ln (k/k_0)$ with $k_0 = 0.05$ Mpc$^{-1}$
\citep{SpergelMAP}. Using these parameters we apply the rescaled ST
approximation to estimate the halo mass function in voids.
Figure~\ref{fig:normal} shows the estimates. As expected, the mass
function is then less steep than the mass functions with the same
normalization and a constant slope $n=1$. On the other hand, decreased
normalization makes it steeper so the net effect is that it is just
shifted down with respect to the mass function found for $\sigma_8=0.9$
and $n=1$.

Note, that in Figure~\ref{fig:normal} the mass function is shown down
to $10^{7}\Msunh$. This is far beyond the range which can be tested in
a numerical experiment, but due to the good agreement between the
numerical simulation and the analytical approach we expect that the
rescaled ST mass function can be extended to smaller masses.  Using
this approximation we predict that in a typical void of diameter $\sim
20\hMpc$ there should be about 100 000 objects of mass greater than
$10^{7}\Msunh$. It is a challenge to observers to confirm this
prediction. Recently, the existence of a large population of hydrogen
clouds in voids has been claimed \citep{Manning2002}. These clouds
could be associated with a fraction of the numerous haloes which we
found in the simulated voids.

\begin{figure}
\includegraphics[width=\columnwidth]{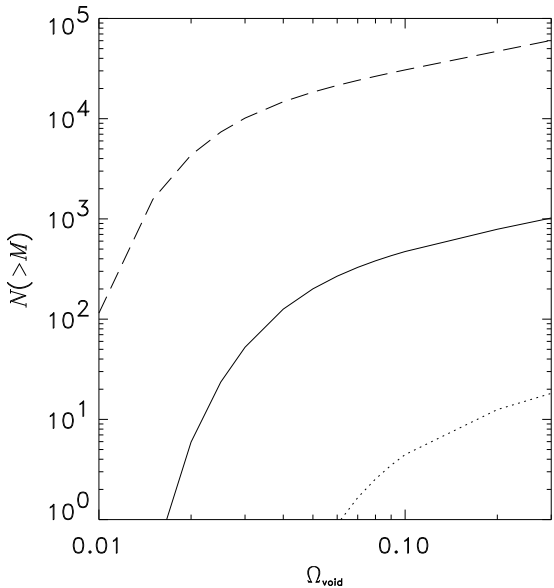}
\caption{The number of haloes with the virial mass larger than
$10^{12}\Msunh$ (bottom curve), $10^{10}\Msunh$ (middle curve), and
$10^{8}\Msunh$ (top curve) inside a typical void of radius $10\Mpch$
as a function of the average density of the void $\Omega_{\rm
void}$. These predictions come from the rescaled unconstrained mass
function. A ``void'' with $\Omega_{\rm void}=0.3$ has the average matter
density in the Universe. }
\label{fig:mass_func_omega1}
\end{figure}

The expected number of dark matter haloes of different mass depends on
the void (under)density. In Figure~\ref{fig:mass_func_omega1} we show
predictions for the number of haloes inside a typical void of radius
$10\Mpch$ for haloes with mass larger than $10^{12}\Msunh$,
$10^{10}\Msunh$ and $10^{8}\Msunh$ as functions of the average density
of the void. There is a dramatic decline in the number of haloes when
the average density of the void falls below a certain percentage of the
mean density. The larger the mass of the haloes, the larger is the void
density at which the sharp decline happens. This decline is related to
the steepening of the void mass function at the high mass end as seen
in Figure~\ref{fig:mass_func_simu_n}. Note, that for typical voids with
the average density ten times below the mean density, the number of
$10^{8}\Msunh$ haloes is not yet suppressed: the sharp decline is at
lower average densities.

\section{Discussion and conclusions}
\label{sec:conclusions}

We used $N$-body simulations to study the formation of voids in the
large-scale structure of the Universe.  We
first identified dark matter haloes and then searched for voids in the
distribution of the haloes. We found that the size of the voids
depends weakly on the lower mass (or circular velocity) limit of the
haloes chosen to define the void. Voids in a set of more massive
haloes tend to be slightly larger than voids in a set of smaller haloes.

The interior of five largest voids was resimulated with very high mass
resolution of $4 \times 10^7 \hMsun$. These voids have diameters of
about $20\Mpch$ and their density is about a factor of 10 smaller than
the mean density in the Universe.  Inside a void of this size we
found typically about 50 haloes with circular velocities larger than
50 km/s and more than 800 with circular velocities larger than 20
km/s. A scale-dependent slope of the primordial power spectrum as
recently suggested by \cite{SpergelMAP} would slightly reduce the
number of low mass haloes in voids.

\citet{Mathis2002} find ``several void regions with diameter $10
\hMpc$ in the simulations where gravity seems to have swept away even
the smallest haloes'' they ``were able to track''.  According to their
Table~1 they tracked haloes down to 10 particles corresponding to halo
masses $3.6 \times 10^{10}\hMsun$. In our simulations of voids of
diameter $20 \hMpc$ we find typically up to 10 haloes of this mass
(represented by 1000 particles). More than half of the haloes are close
to the outer void boundary  with a distance to the void's centre larger than
80\% of the void's radius. Since our void volume is 8 times larger
and more massive haloes tend to be situated in the outer part of the
void it would be not extremely unprobable to find inside our voids
regions of $10 \hMpc$ free of any halo more massive than $3.6  \times
10^{10} \hMsun$. In this respect we agree with
\cite{Mathis2002}. Following the evolution of voids numerically it seems
that these haloes are not swept away by gravity but never form in
regions with the lowest density.

The fact that regions devoid of haloes with masses larger than $5 \times
10^{10}\hMsun$ are smaller than those devoid of haloes with masses bigger
than $2 \times 10^{11}\hMsun$ emphasizes our argument that the size of the
voids depends on the objects used to define the void. For example, the
voids in the distribution of clusters are not empty: they contain many
$L_*$ galaxies. Likewise, the voids in a sample of $L_*$ galaxies
should contain dwarfs.  This argument suggests that voids in the dark
matter distribution should be self-similar in the same sense as
clusters of galaxies have hundreds of galaxies and galaxies have
hundreds of satellites. It must be so as long as only the gravity is
the dominant factor and the spectrum of fluctuations is approximately
scale invariant. The question whether luminous galaxies form or do not
form in all these dark matter haloes is then the most important
question in the theory of galaxy formation in different environments.

Assuming with \cite{Mathis2002} a luminosity $M_B = -16.5$ for a galaxy
hosted by a halo of $3.6 \times 10^{10}\hMsun$ we predict about five of
these galaxies to be found in the inner part of a typical void of
diameter $20 \hMpc$.  In principle they can be detected. In the giant
(radius 31.5 Mpc) Bo\"otes void \citet{Szomoru1996a} study void
galaxies in about 1\% of the void volume.  All but one of them were
substantially brighter than $M=-16.5$. The one reported with $M=-16.2$
is the companion of a brighter one. To compare the model predictions
with observations one would have to study a void in the distribution of
galaxies with limiting magnitude $M_B$ between $-17.5$ and $-18.5$ which
roughly corresponds to our threshold mass of $2 \times 10^{11} \hMsun$.

Note, that so far we assumed that each dark matter halo hosts a
galaxy. This may not be true. Physical processes of galaxy formation
are not well known and there could be processes that strongly
suppress the formation of stars inside small haloes, which collapse
relatively late in voids. One such process which has been widely discussed
is the ionizing flux (e.g. \citealt{Bullock00,Bensonetal2002}).

We determined the mass function of dark matter haloes in the simulated
voids and compared it with the analytical predictions based on the
Sheth-Tormen formalism. The formalism was applied in two versions: (1) We
used the ansatz for the constrained mass function proposed by
\cite{Sheth2002} and (2) we proposed our own extension of the unconstrained
mass function of \cite{Sheth1999} by rescaling the power spectrum in
the void. We found that our approach is not only much simpler in
application, but also reproduces more accurately the mass functions
obtained in the simulations of voids.

\section*{Acknowledgments}

S.G., A.K. and Y.H. acknowledge support by NATO grant PST.LLG.978477,
A.K. and S.G. by NSF/DAAD, E{\L} by the Polish KBN grant No.  2P03A00425.
Research was supported by NASA and NSF grants to NMSU,
YH by the Israel Science Foundation grant 143/02. 
E{\L}, AK and YH are grateful for the hospitality of Astrophysikalisches
Institut Potsdam where most of this work was done. Computer simulations
presented in this paper were done at the Leibnizrechenzentrum (LRZ) in
Munich and at the National Center for Supercomputer Applications (NCSA).

\label{lastpage}

\end{document}